\documentclass{article}

    \PassOptionsToPackage{numbers, compress}{natbib}
 \usepackage[dblblindworkshop,final]{neurips_2025}
\usepackage{amssymb}

\usepackage[utf8]{inputenc} % allow utf-8 input
\usepackage[T1]{fontenc}    % use 8-bit T1 fonts
\usepackage{hyperref}       % hyperlinks
\usepackage{url}            % simple URL typesetting
\usepackage{booktabs}       % professional-quality tables
\usepackage{amsfonts}       % blackboard math symbols
\usepackage{nicefrac}       % compact symbols for 1/2, etc.
\usepackage{microtype}      % microtypography
\usepackage{xcolor}         % colors
\usepackage{amsmath} 
\usepackage{paralist}
\usepackage{graphicx}
\usepackage{tabularx} 
\title{Continuous-Token Diffusion for Speaker-Referenced TTS in Multimodal LLMs}

\workshoptitle{Structured Probabilistic Inference \& Generative Modeling (SPIGM)}

\author{%
\begin{tabular}{@{}c@{}}
Xinlu He$^{1}$, Swayambhu Nath Ray$^{2}$, Harish Mallidi$^{2}$, Jia-Hong Huang$^{2}$,\\ Ashwin Bellur$^{2}$,
Chander Chandak$^{2}$, M. Maruf$^{2}$, Venkatesh Ravichandran$^{2}$
\end{tabular}\\[3pt]
$^{1}$Worcester Polytechnic Institute, USA \quad $^{2}$Amazon AGI, USA
}

\begin{document}

\maketitle

\begin{abstract}
Unified architectures in multimodal large language models (MLLM) have shown promise in handling diverse tasks within a single framework. In the text-to-speech (TTS) task, current MLLM-based approaches rely on discrete token representations, which disregard the inherently continuous nature of speech and can lead to loss of fine-grained acoustic information.
In this work, we investigate the TTS within the MLLM paradigm using continuous speech representations.
We design a dual-head architecture and implement two complementary training strategies for a robust model. (1) A diffusion head generating continuous speech representations is added on the MLLM, which is on frame-level and strictly autoregressive. (2) The original language model head is retained to preserve multitask capability and to control the start and end of speech synthesis.
(3) Masked training is employed to address exposure bias in autoregressive decoding.
(4) To stabilize optimization, we propose a two-stage scheme where the LM is frozen in the second stage, ensuring the diffusion head learns from a fixed input distribution. Evaluations on LibriSpeech(PC) test-clean show that our approach achieves state-of-the-art autoregressive performance, with a WER of 1.95\%, speaker similarity of 0.54, and UTMOS of 4.00. The two-stage training yields a 46\% relative WER reduction over the one-stage training baseline. These results highlight the effectiveness of combining autoregressive modeling with continuous-token diffusion, supported by a two-stage training procedure.
\end{abstract}

\section{Introduction}
\label{sec:intro}
Recent advances in multimodal large language models (MLLMs) have enabled a single model to perform diverse tasks across modalities in an autoregressive manner \cite{Alayrac2022Flamingo, trinh2024DMLM, mllm_Yin_2024}. In text-to-speech (TTS), the dominant approach converts speech into discrete tokens \cite{guo2023xtts,valle}, allowing TTS to be framed as a sequence prediction problem within the LLM framework. While effective, discrete quantization can discard fine-grained acoustic details, limiting naturalness and speech fidelity. 
In contrast, continuous speech representations—often learned by variational autoencoders (VAEs) or other self-supervised encoders \cite{baevski2020wav2vec20frameworkselfsupervised,2019spl,hsu2017speechvae}—better preserve the intrinsic properties of speech. Diffusion models, originally successful in high-fidelity image generation \cite{DDPM, DDIM, igdiffsurvey}, have recently achieved state-of-the-art results for TTS by modeling such continuous representations in a non-autoregressive manner \cite{ proddiff, mehta2024matcha,Popov2021GradTTSAD}. However, the integration of continuous-token diffusion into an autoregressive MLLM framework remains largely unexplored.

\begin{figure}[!t]
\centering
\includegraphics[width =\linewidth ]{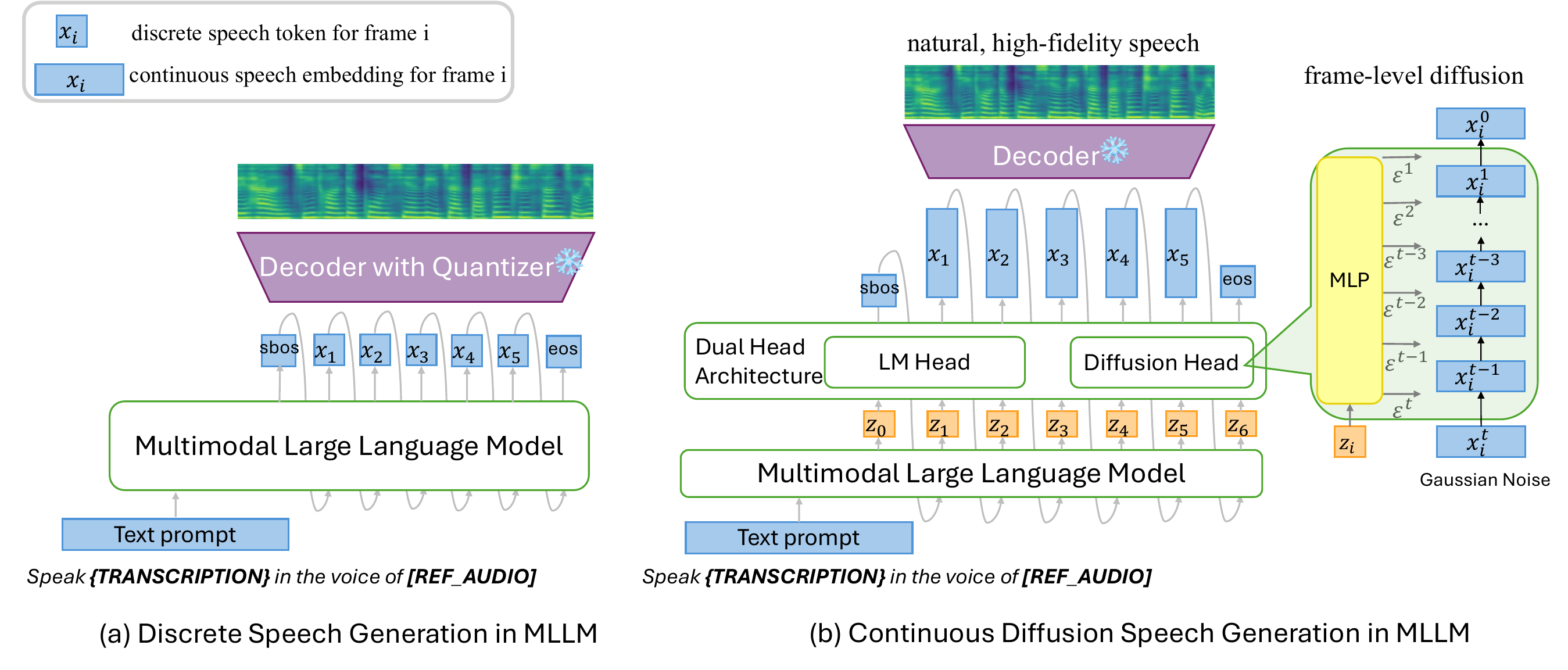}
\caption{ \textbf{Illustration of the proposed method}. Compared to discrete token-based generation in (a), our approach (b) adopts a dual-head MLLM with a diffusion head for frame-level autoregressive generation, enabling continuous speech synthesis with natural and high-fidelity quality.
% \textcolor{red}{\textcolor{blue}{JH:} (do this figure as soon as possible, or it's a bit difficult to understand the whole method and then refine your introduction.)}
}
\label{fig:method}
\end{figure}

Building on this line, we propose to integrate diffusion directly into an autoregressive MLLM framework, as shown in Fig.~\ref{fig:method}. Prior autoregressive diffusion methods either rely on intermediate semantic tokens \cite{SALADS_IBM}, or relax framewise causality in diffusion modules to predict multi-frame blocks per AR step \cite{ DiTAR,liu2024autoregressivediffusiontransformertexttospeech}. In contrast, our approach implements a strictly frame-by-frame autoregressive, continuous-representation diffusion head on top of an LLM backbone. This design enables the model to directly generate high-fidelity speech in the continuous speech representation space, avoiding the quantization bottleneck. 
% To address these limitations, we implement a fully autoregressive, frame-level, continuous-token diffusion head on top of an LLM backbone. This design enables the model to directly generate high-fidelity speech in the continuous representation space, avoiding the quantization bottleneck. 

To maintain the MLLM’s multi-task consistency, we designed a dual-head architecture. The diffusion head generates continuous speech embeddings each frame, which are then decoded to synthesize waveforms. The language model (LM) head retained from the original LLM predicts the start and end of speech tokens, enabling variable-length speech synthesis. This token-based control enables seamless integration of speech generation with multimodal generation. Unlike prior TTS methods \cite{li2019neuralspeechsynthesistransformer, meng2025autoregressivespeechsynthesisvector} introducing an external classifier module, our design preserves a single, unified framework within the MLLM.

Our investigation into continuous-token autoregressive TTS revealed two major training challenges: exposure bias and joint optimization instability. First, the mismatch between teacher-forced training and free-running inference causes small frame-level deviations to accumulate over long sequences. We mitigate this exposure bias \cite{bengio2015scheduled} by introducing a masked training scheme inspired by audio generation \cite{yang2025generativeaudiolanguagemodeling}, where a proportion of ground-truth frames is masked during training. This bridges the gap between training and inference conditions, improving robustness and temporal consistency in generated speech.

Then we observed that jointly optimizing the diffusion head with the LLM is unstable. The LLM’s output distribution evolves during training, causing instability in the diffusion learning process. To address this, we employ a two-stage training strategy: in stage 1, train the LLM and diffusion head jointly, allowing the LLM to learn speech token prediction and the diffusion head to adapt to evolving inputs. In stage 2, we freeze the entire LLM side, including the backbone, the LM head, and the speech-projection, thereby fixing the input distribution to the diffusion head. We then train only the diffusion head. This allows the diffusion model to focus solely on refining the mapping from the LLM outputs to the target speech space.
This separation stabilizes training and significantly improves generation quality.

Our main contributions are as follows:
\begin{compactitem}
% \begin{itemize}
    \item We introduce a frame-by-frame continuous-token diffusion head into an autoregressive MLLM for speaker-referenced TTS, distinguishing it from block-wise multi-frame designs.
    % We implement a fully autoregressive, frame-level, continuous-token TTS framework, and examine the applicability of the diffusion head in MAR and AudioNTP to speech synthesis.
    \item We propose a dual-head architecture where LM head supports variable-length speech and keeps unified multimodal framework.
    \item We mitigate autoregressive exposure bias via masked training, improving temporal consistency and model robustness.
    % We propose a dual-head architecture in which the language model head generates special \texttt{<bos>} and \texttt{<eos>} tokens for speech, replacing the use of a separate end-of-sequence classifier. We demonstrate the effectiveness of this approach and show that it is compatible with unified multi-task foundation model frameworks.
    % We analyze the exposure bias problem in autoregressive speech generation and introduce a masking strategy to mitigate error accumulation during generation.
    \item We stabilize training with a two-stage strategy, yielding large performance gains and cutting WER by 46\%, reaching SOTA AR on LibriSpeech(PC) test-clean.\footnote{The models and results described in this paper are intended for research purposes only.}
    % We further investigate training strategies for autoregressive diffusion models. Given the longer training requirements of diffusion, we propose a two-stage training scheme to improve efficiency and stability.
% \end{itemize}
\end{compactitem}
\section{Related Work}
\label{sec:related_work}
\textbf{Zero-shot TTS.} 
Zero-shot text-to-speech (TTS) refers to synthesizing speech for previously unseen speakers by leveraging a short reference utterance as conditioning, thereby enabling speaker generalization without explicit speaker-specific training. 
Inspired by advances in large language models (LLMs)~\cite{brown2020languagemodelsfewshotlearners}, zero-shot TTS is often formulated as a language modeling task~\cite{borsos2023audiolmlanguagemodelingapproach,zhang2023speechgptempoweringlargelanguage}, where speech waveforms are transformed into sequences of tokens and synthesized via next-token prediction. 
Existing methods can be broadly categorized into multi-stage and single-stage pipelines. 
Multi-stage systems, such as VALL-E~\cite{valle} and SALAD~\cite{SALADS_IBM}, autoregressively predict coarse units such as semantic~\cite{borsos2023audiolmlanguagemodelingapproach} or codec tokens~\cite{zeghidour2021soundstreamendtoendneuralaudio}, which are then refined into waveforms. 
This decomposition improves stability but often discards fine-grained acoustic details. 
In contrast, single-stage approaches, exemplified by MegaTTS~\cite{megatts} and NaturalSpeech~\cite{ju2024naturalspeech3zeroshotspeech}, directly generate high-information continuous representations, offering higher fidelity while facing greater challenges in robustness. 
Our method follows this single-stage paradigm. 

\textbf{Autoregressive Diffusion.} 
Autoregressive language models were originally developed for discrete symbol sequences, whereas diffusion models are particularly effective for continuous data distributions~\cite{DDPM}. 
Recent work has explored combining these paradigms for sequence generation. 
Several studies modify the diffusion process to behave autoregressively, for example by adjusting denoising schedules so earlier tokens are predicted before later ones~\cite{chen2023analogbitsgeneratingdiscrete,fang2024vectorquantizeddiffusionmodel}.
TransFusion~\cite{zhou2025transfusion} exemplifies this strategy with a shared transformer that applies causal attention to discrete tokens and bidirectional attention to continuous features, though it still struggles with strictly causal generation of continuous signals. 
Other efforts replace discrete codec units with continuous-valued tokens modeled directly through diffusion losses~\cite{li2024MAR,yang2025generativeaudiolanguagemodeling}. 
By avoiding quantization, these approaches preserve fine-grained semantic and acoustic detail, positioning diffusion as a compelling alternative to conventional autoregressive modeling. 

\section{Proposed Method}
\label{sec:method}
We aim to autoregressively generate speech in the space of continuous acoustic embeddings. Our framework builds upon a large language model backbone 
% to perform frame-level autoregressive speech generation from continuous acoustic representations. We 
and introduces a dual-head architecture. The first part of the method addresses how continuous speech tokens are generated: Section~\ref{sec:diffusion} introduces continuous-token generation with a diffusion head, and Section~\ref{sec:dualhead} presents EOS control in the dual-head architecture, which together form the foundation for continuous speech generation within a multitask unified foundation model setting. The second part focuses on improving robustness and overall performance: Section~\ref{sec:mask} describes masked autoregressive learning, which exposes the model to imperfect histories, and Section~\ref{subsec:two_stage_training} details a two-stage optimization scheme, which stabilizes training by mitigating distribution drift.
\subsection{Continuous-token Generation with Diffusion Head}
\label{sec:diffusion}
With the inherent continuous nature of speech, recent work has begun to adopt continuous representation for speech generation in TTS. Compared to discrete codebook tokens, continuous representations preserve fine-grained characteristics, while decreasing the potential information loss~\cite{li2025continuousspeechtokenizertext}. Inspired by image~\cite{li2024MAR} and audio~\cite{yang2025generativeaudiolanguagemodeling} generation, we introduce a lightweight diffusion head on top of a causal foundation model to generate high-fidelity speech from continuous embeddings.

Formally, the target is a sequence of continuous speech embeddings at frame level $x = \{x_1,\ldots,x_N\}$, where $x_i\in\mathbb{R}^d$ denotes the speech embedding in frame $i$. An off-the-shelf variational autoencoder $V=\{V_E,V_D\}$ provides the waveform-embedding mapping: the encoder $V_E$ extracts embeddings from waveform $w$, and the decoder $V_D$ reconstructs audio from predicted embedding $\hat{x}$. 

\textbf{Inference.} Fig.~\ref{fig:method} (b) illustrates the framework. 
Given a prompt $p$ with transcription and reference audio, multi-modal causal LLM $\mathcal{C_{\theta}}$ takes $p$ and past predictions $\hat{x}_{<i}$ to autoregressively produce a hidden state as condition $z_i=\mathcal{C}_\theta(p,\hat{x}_{<i})$. This vector $z_i$ conditions a diffusion head with MLP denoiser $M_\phi$, which starts from Gaussian noise $x_i^t\sim\mathcal{N}(0,I)$ and iteratively denoises to produce the next embedding $\hat{x}_i=x_i^0$.

\textbf{Training.} During training,
% we form a noised target samples $X_i^T\sim\mathcal{N}(0,\mathbf{I})$, and iteratively applies the learned denoising updates from $M_{\phi}$ to get predicted embedding $x_i^0$. During diffusion traininggiven the ground-truth frame embedding $x_i$, 
we sample a total timestep $t\!\sim\!\mathit{U}\{1,\dots,T\}$ for adding noise and noise $\varepsilon\!\sim\!\mathcal{N}(0,I)$, and form a noised target $x_i^t=\sqrt{\bar{\alpha}_t}x_i+\sqrt{1-\bar{\alpha}_t}\varepsilon$, where $\bar{\alpha}_t$ defines a noise schedule \cite{DDPM,nichol2021improveddenoisingdiffusionprobabilistic}.
A small MLP denoiser $M_\phi$ predicts the noise $\hat\varepsilon=M_\phi\!\big(x_i^t,\, t,\, z_i\big)$, where $x_i^t,t,z_i$ denotes the current state, the timestep, and the LLM condition $z_i$.  We minimize the standard noise-prediction loss
\[
\mathcal{L}_{\text{diff}}(\theta,\phi)
\;=\;
\mathbb{E}_{t}\!\left[
\left\|
\varepsilon -\hat\varepsilon
\right\|^2
\right],
\]
which backpropagates through $z_i$. This diffusion loss can be jointly optimized with the objective of the language model head. Further details are provided in the following sections.

\subsection{EOS Control in Dual-Head Architecture}
\label{sec:dualhead}
Prior speech generation approaches~\cite{valle} typically rely on an auxiliary classifier or a fixed-length constraint to determine the endpoint of speech output. In contrast, our framework delegates boundary control to the LM head, enabling seamless integration into a unified multi-task backbone. 
 
\textbf{Inference.} Generation proceeds under the control of the LM head. The model begins in the textual phase until the LM head emits a special token \textit{<speech\_bos>}, which triggers the speech generation phase. 
At each subsequent step, the LM head produces a control token. If it emits \textit{<cont\_speech\_gen>}, this token is not added to the output sequence; instead, it signals the diffusion head to generate the next speech embeddings. Otherwise when LM head predicts \textit{<eos>}, the speech generation phase terminates.
This token-based mechanism provides a unified and modality-agnostic interface for switching between text and speech without additional architectural components.

\textbf{Training.} As shown in Fig~\ref{fig:res}, in order to supervise this control process, we extend the vocabulary with a special token \textit{<cont\_speech\_gen>} in addition to \textit{<speech\_bos>} and \textit{<eos>}. Although \textit{<cont\_speech\_gen>} is not emitted during inference, its explicit supervision during training provides dense learning signals throughout the speech segment. This design reduces the risk of the LM head prematurely predicting \textit{<eos>} in variable-length sequences, compared with supervising only the boundary tokens. The LM head is trained with the standard cross-entropy $\mathcal{L}_{\text{LM}}$ over the control tokens, while the diffusion head is trained with the noise-prediction loss $\mathcal{L}_{\text{diff}}$ (Sec.~\ref{sec:diffusion}). The overall objective is the sum of the LM cross-entropy loss and the diffusion loss:
\[
\mathcal{L} = \mathcal{L}_{\text{LM}} + \mathcal{L}_{\text{diff}}
\]

\subsection{Masked Autoregressive Learning}
\label{sec:mask}
\begin{figure}[!t]
\centering
\includegraphics[width =\linewidth ]{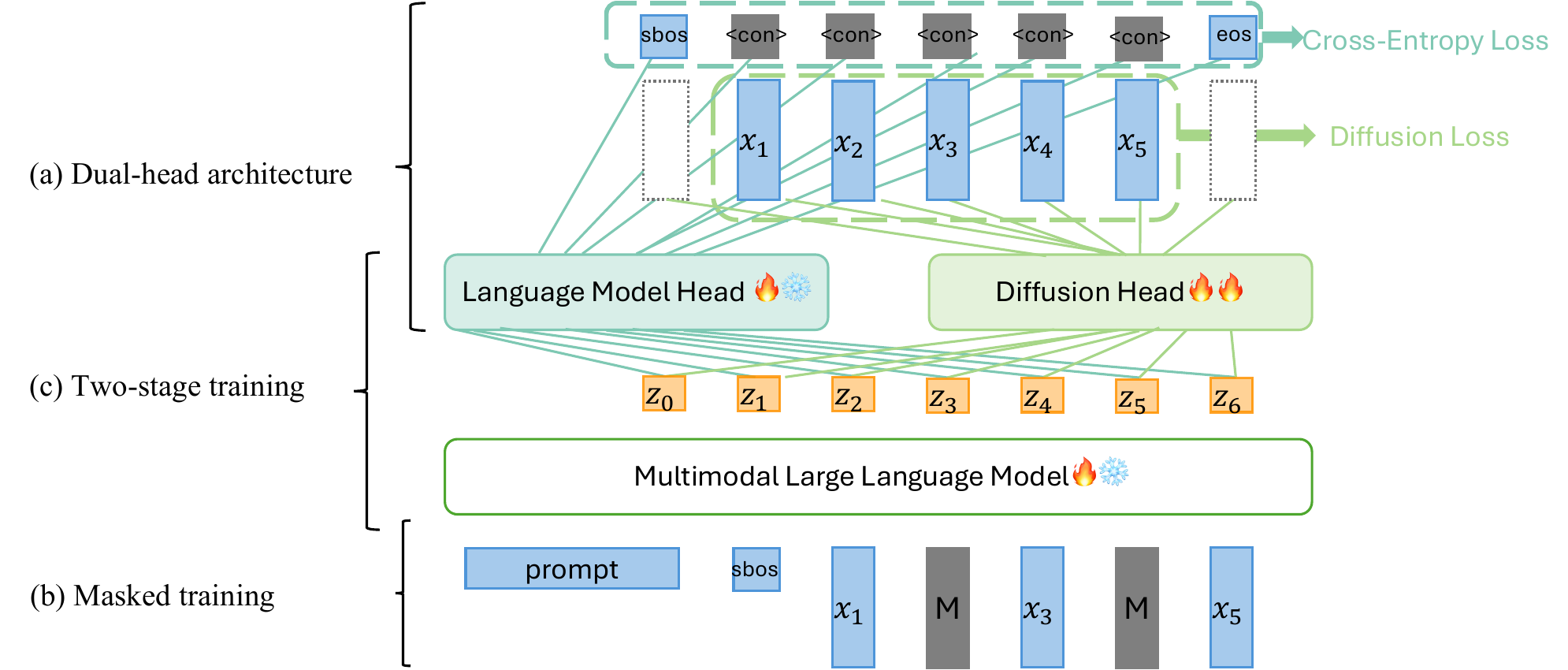}
\caption{\textbf{Training and design details of our method.} (a) Dual-head architecture: The LM head predicts speech boundaries (<s-bos> and <eos>) and the special token <continuous\_speech\_gen>, while the diffusion head generates continuous frame-level speech embeddings. (b) Masked training strategy: A portion of speech inputs is masked to bridge the gap between teacher-forcing training and auto-regressive inference. (c) Two-stage training: In the first stage, all components are trained jointly; in the second stage, only the Diffusion head is further optimized.}
\label{fig:res}
\end{figure}
Autoregressive generation is affected by exposure bias \cite{bengio2015scheduled}, where models are trained on ground-truth histories but must rely on its own potentially erroneous predictions during inference, resulting in error accumulation. 
To mitigate this issue, various strategies such as masking have been explored in text generation~\cite{ghazvininejad2019maskpredictparalleldecodingconditional} and audio generation~\cite{yang2025generativeaudiolanguagemodeling}. Motivated by these advances, we adapt masking to continuous frame-level speech generation.
 
During training, before feeding the acoustic embedding sequence $x = \{x^1, \ldots, x^N\}$ into the causal predictor $C_\theta$, we apply \textit{zero embedding masking} to simulate imperfect histories. We define a binary mask $v = \{v^1, \ldots, v^N\}, \; v^t \in \{0,1\}$, where each entry is sampled independently as $v^t \sim \mathrm{Bernoulli}(1-p_{\mathrm{mask}})$. Thus, with probability $p_{\mathrm{mask}}$, the corresponding frame is masked and replaced by the zero vector. The corrupted sequence is then $\tilde{x} = x \odot v$, where $\odot$ denotes element-wise multiplication, so masked positions are replaced by the zero vector.
This zero embedding masking strategy can be viewed as input-level masking, where the masking ratio $p_{\mathrm{mask}}$ explicitly controls the level of corruption in the autoregressive history. The training masking is shown in Fig.~\ref{fig:res} (b).
At inference time, no masking is applied. The model operates autoregressively on its own predictions. 
By training under such corrupted contexts, the model is encouraged to handle imperfect histories more robustly, thereby mitigating exposure bias and improving stability in long-form speech synthesis.  
\subsection{Two-Stage Scheme}
\label{subsec:two_stage_training}

When training the MLLM and diffusion head jointly in an end-to-end manner, we observed instability caused by distribution drift. The MLLM output distribution $p_\theta(z \mid p)$ evolves as the parameters $\theta$ are updated. The diffusion head $\mathcal{M}_\phi$ is expected to learn a mapping from $z$ to continuous speech embeddings $x$. However, since the source distribution $p_\theta(z)$ is non-stationary during training, $\mathcal{M}_\phi$ must adapt to a shifting input space, making convergence unreliable and degrading generation quality. We hypothesize that freezing $\theta$ to fix $p_\theta(z)$ yields a stationary input distribution, allowing the diffusion head $\mathcal{M}_\phi$ to focus on modeling a stable transformation, thereby improving optimization stability and synthesis fidelity.

To implement this idea, we adopt a two-stage training strategy, as shown in Fig.~\ref{fig:res}(c). In \textbf{Stage 1}, we jointly train the causal LM $\mathcal{C}\theta$ and the diffusion head $\mathcal{M}\phi$ in an end-to-end manner by minimizing the sum of the cross-entropy loss and the diffusion loss. This stage enables the model to align the LM outputs with the target distribution and to produce coarse speech embeddings. However, we observe that although the overall training objective consistently decreases, autoregressive evaluation metrics exhibit a non-monotonic trend, improving initially but then deteriorating. This is consistent with our hypothesis that distribution drift hinders stable refinement. 

In \textbf{Stage 2}, we freeze the MLLM and LM head parameters $\theta$ and train only the diffusion$\mathcal{M}\phi$. With $p\theta(z)$ fixed, the input distribution to the diffusion head remains stationary, allowing it to focus on refining LM outputs into high-fidelity acoustic frames.
This stage stabilizes optimization, mitigates the instability observed in joint training, and leads to improved synthesis quality.

\section{Experiments}
\label{sec:exp}
To evaluate our dual-head continuous speech generation framework, we conduct TTS experiments under different settings, comparing intelligibility, speaker identity preservation, and speech naturalness. We first introduce the datasets and evaluation protocols in Sec.~\ref{subsec:dataset_metric}, then present implementation details in Sec.~\ref{subsec:implementation}, and finally describe the baselines in Sec.~\ref{subsec:baseline}.

\subsection{Dataset and Metrics}
\label{subsec:dataset_metric}
\textbf{Datasets} We adopt the LibriVox corpus as the training source. Specifically, we use a 50k-hour subset (derived from the Libri-Light collection~\cite{librilight}), which consists of read English audiobooks from thousands of speakers. For evaluation, we use the Librispeech (PC) test-clean dataset. Following the protocol of NaturalSpeech~\cite{ju2024naturalspeech3zeroshotspeech}, we randomly select one utterance from each of 40 speakers and use an additional 3-second clip as the speaker reference. 

\textbf{Evaluation metrics}

We evaluate our system on three aspects: intelligibility, speaker identity preservation, and speech quality. 
(1) Word error rate (WER) measures intelligibility, about how accurately the synthesized speech conveys the reference text. The generated speech is transcribed by Whisper-Large~\cite{radford2022whisper}, and WER is calculated from insertions, substitutions, and deletions. 
(2) Speaker similarity is measured as cosine similarity between embeddings extracted by ECAPA-TDNN~\cite{desplanques2020ecapa}. We report SIM-R (to reference prompt) and SIM-G (to ground-truth speech). 
(3) Speech quality is estimated with UTMOS~\cite{UTMOS}, an objective MOS predictor trained on large-scale human ratings. 

\subsection{Implementation Details}
\label{subsec:implementation}
\textbf{Speech Representation.}
Our system relies on two types of speech representations: one for speaker reference prompting and another for generation targets.
For speaker reference prompting, we extract a 768-dimensional embedding from a three-second reference clip using LAM~\cite{male25_interspeech_LAM}. This embedding encodes speaker identity and is projected into the LLM input space as conditioning information.

For generation targets, we adopt a pretrained VAE-based vocoder that encodes speech into 64-dimensional embeddings at 25 frames per second.
Given a stereo waveform sampled at 48\,kHz, the encoder applies a pre-convolutional projection to 128 channels, followed by five downsampling ResNet blocks, producing frame-level features at 25\,Hz. A post-convolutional projection and VAE bottleneck is then applied to yield frame-level 64-dimensional latent vectors.

\textbf{Model Details}
Our architecture builds on an autoregressive LLM backbone and extends it with projection modules for multi modality and a diffusion head for speech generation. We adopt OPT-125M~\cite{opt} as the LLM backbone. For multi-modality, we add two projectors: one maps the 768-dimensional reference embedding extracted from the speaker prompt into the LLM input space, and the other maps the 64-dimensional continuous acoustic tokens into the LLM input space. With these projections, the backbone functions as a multimodal language model (MLLM) rather than a purely text-based LLM.

We use the final hidden state from the LLM decoder as the input to the diffusion module. Before entering the diffusion process, this hidden representation is projected through the third linear layer to 768 dimensions as the diffusion condition. The diffusion head is implemented as a stack of MLP layers and operates with a DDPM-based denoising process, similar to ~\cite{li2024MAR,yang2025generativeaudiolanguagemodeling}.

\textbf{Diffusion Hyperparameters}
During training, we use a diffusion process with $T=1000$ steps and adopt the cosine noise schedule~\cite{nichol2021improveddenoisingdiffusionprobabilistic},where derives $\beta_t$ implicitly from the cumulative product $\bar{\alpha}_t$.
The diffusion head is an MLP with residual blocks; we experimented with 3, 6, and 12 layers, and report 12-layer results unless otherwise noted. 
Each block consists of layer normalization, linear layers, and SiLU activation with adaptive layer normalization modulation, with no dropout. 
During inference, we reduce the denoising process to 100 steps and apply a sampling temperature of $0.9$. CFG is set to 1.

\textbf{Training Configuration.} 
All experiments are conducted on NVIDIA A100 GPUs with a global batch size of 2048. 
We use the Adam optimizer without weight decay and employ FP16 mixed precision for efficiency. 
In stage~1, the learning rate is linearly warmed up from $3\times10^{-5}$ to $3\times10^{-4}$ over the first steps and then decayed to zero using a cosine schedule, for a total of 300k steps. 
In stage~2, the model is further trained for 300k steps with a constant learning rate of $2\times10^{-4}$.

\subsection{Baseline}
\label{subsec:baseline}
We use the model from the first-stage joint training of all components, including the MLLM backbone, LM head and diffusion head, as our baseline. During training, the loss decreases monotonically, but the evaluation WER first decreases and then increases because of the dynamic condition for the diffusion head. We therefore apply the early stopping and select the checkpoint with the lowest validation WER as the baseline model, from which the stage-2 training is initialized.

\section{Results}
We first present the main results by comparing our model with representative baselines. 
We then provide ablations and analyses of key design choices and inference settings to understand their impact on intelligibility, speaker similarity, and naturalness.
\label{sec:results}
\subsection{Main Results}
\begin{table*}[!t]
\centering
\caption{Objective evaluation on LibriSpeech(PC) test-clean. Results are reported for WER (\%), speaker similarity (cosine SIM), and UTMOS. $^\dagger$ denotes results reproduced by NaturalSpeech3. Our method outperforms larger hybrid AR+NAR baselines while using fewer parameters.}
\resizebox{\linewidth}{!}{%
\begin{tabular}{lcccccc}
\toprule
\textbf{Method} & \textbf{Modeling} & \textbf{Token} & \textbf{\# Params} & \textbf{WER(\%)$\downarrow$} & \textbf{SIM}$\uparrow$& \textbf{UTMOS$\uparrow$}  \\
\midrule
Ground Truth& - &- &- & 2.84 & 0.69 & 4.16  \\
Vocoder&- &- &- &  2.56 & 0.61 & 3.82 \\
\midrule
VALL-E$^{\dagger}$ (~\cite{valle})& AR+NAR & Discrete & 400M &  6.11 & 0.47 & 3.68 \\
Mega TTS$^{\dagger}$ (~\cite{megatts}) & AR+NAR & Continuous & 500M  & 2.32  & 0.53& \textbf{4.02} \\
Voicebox$^{\dagger}$ (~\cite{le2023voiceboxtextguidedmultilingualuniversal}) & NAR & Continuous &400M   & 2.14  & 0.48& 3.73 \\
StyleTTS2$^{\dagger}$ (~\cite{li2023styletts2humanleveltexttospeech}) & NAR & Continuous &  700M & 2.49  & 0.38& 3.94  \\
\midrule
% RVQ & AR  & Disc  &  &LrbriVox & &   &   & \\
% Continuous MSE & AR  & Cont  & 130M &LrbriVox & &   &   & \\
Stage-1 Baseline & AR  & Continuous  & 160M &3.61&0.49&   3.21 \\
% Proposed Methods& AR  & Cont  & 160M &LrbriVox &1.65&0.52 &  3.76  \\
Proposed Method & AR  & Continuous  & 160M &\textbf{1.95}&\textbf{0.54} &  4.00  \\
\bottomrule
\end{tabular}
\label{tab:sot-comp}
}
\end{table*}
Table~\ref{tab:sot-comp} reports the comparison between our model and representative hybrid autoregressive baselines. 
VALL-E, which relies on discrete tokens, yields a WER of 6.11\% and a speaker similarity of 0.47, illustrating the limitations of quantization in preserving fine-grained acoustic details. 
MegaTTS, a continuous-token model with 500M parameters, achieves stronger results with a WER of 2.32\% and a similarity of 0.53. 
By contrast, our model attains a WER of 1.95\% and a similarity of 0.54, while maintaining competitive perceptual quality (UTMOS 4.00 compared with 4.02 for MegaTTS). 
Despite using only 160M parameters, our system consistently outperforms larger models, demonstrating the efficiency of combining an autoregressive backbone with a continuous diffusion head. 
For additional context, we also include ground-truth speech and vocoder reconstructions in the table for reference.  

We further analyze the effect of our two-stage training strategy. 
The Stage-1 baseline achieves a WER of 3.61\%, a speaker similarity of 0.49, and a UTMOS of 3.21, indicating that the diffusion head initially suffers from unstable input distributions. 
After Stage-2 training, the WER is reduced by 46\% relative (from 3.61\% to 1.95\%), while speaker similarity increases from 0.49 to 0.54 and UTMOS rises from 3.21 to 4.00. 
These improvements show that stabilizing the diffusion head’s input distribution in Stage-2 not only reduces recognition errors but also enhances both speaker consistency and perceived naturalness.  

\subsection{Ablation and Analysis}
To better understand the contributions of different components and design choices, we conduct a series of ablation and analysis experiments. 
We first investigate the effect of masked training, which aims to mitigate exposure bias during autoregressive decoding. 
We then examine the role of diffusion head capacity and the impact of our two-stage training strategy. 
Finally, we analyze the influence of stopping criteria and inference hyperparameters, highlighting how these factors jointly affect intelligibility, speaker similarity, and naturalness. 

\textbf{Masked Ratio.}
Table~\ref{tab: dropout_rate} shows the impact of different masking rates in the masked training scheme. Without masking (0\%), the model suffers from severe exposure bias, leading to a WER of 15.06\%. Introducing moderate masking improves robustness, with the best performance at 30\% masking (WER 6.17\%, UTMOS 3.21). However, excessive masking (50\%) degrades both intelligibility and naturalness, as too much corruption disrupts semantic alignment. This confirms that moderate masking helps bridge the gap between training and inference conditions.
\begin{table}[h]
% \begin{minipage}{0.48\linewidth}
\centering
\caption{Performance with different masking rates, using a 3-layer MLP diffusion head. }
% \resizebox{0.8\linewidth}{!}{%
\begin{tabular}{cccccccccc}
\toprule
\textbf{Masking Rate(\%)} & \textbf{WER (\%)$\downarrow$}& \textbf{SIM-R}$\uparrow$&  \textbf{SIM-G}$\uparrow$ & \textbf{UTMOS} $\uparrow$ \\
\midrule
0& 15.06 & 0.45& 0.42 & 2.00  \\
15&12.65 & 0.45 & 0.42 & 1.39  \\
30& 6.17& 0.46 &0.43  & 3.21 \\
50& 8.13  & 0.46  & 0.43 &2.84 \\
\bottomrule
\end{tabular}
\label{tab: dropout_rate}
% }
\end{table}

\textbf{Diffusion Head Depth.}
Table~\ref{tab: diffusion_layer} compares different diffusion head depths with and without the proposed two-stage training. Increasing the depth from 3 to 12 layers progressively improves WER and speaker similarity, demonstrating the benefit of a stronger decoder. More importantly, enabling two-stage training further reduces WER to 1.95\% and boosts similarity and naturalness, highlighting the effectiveness of stabilizing the diffusion head with a fixed input distribution.
\begin{table}[h]
\centering
\caption{Comparison of different numbers of MLP layers in the diffusion head, with dropout rate set to 30\%. Stage-2 FT indicates whether two-stage fine-tuning is applied.}
% \resizebox{\linewidth}{!}{%
\begin{tabular}{ccccccccc}
\toprule
\textbf{\# MLP} & \textbf{Stage-2 FT} & \textbf{\# Params} & \textbf{WER (\%)$\downarrow$} & \textbf{SIM-R}$\uparrow$&  \textbf{SIM-G}$\uparrow$ & \textbf{UTMOS} $\uparrow$ \\
\midrule
3 & w/o&148.7M & 6.17 &0.46& 0.43 & 3.10  \\
6& w/o&164.4M & 5.12 &0.50 & 0.46 & 3.10 \\
12&w/o &159.9M&3.61&0.49 &0.46   & 3.21 \\
12& w&159.9M&1.95&0.54 &0.50&  4.00 \\
\bottomrule
\end{tabular}
% }
\label{tab: diffusion_layer}
\end{table}

\textbf{Stopping Criteria.}
Table~\ref{tab:end_point} compares different stopping strategies. Using ground-truth durations causes unstable outputs (WER 29.36\%), while oracle endpoint supervision achieves low WER but requires non-causal labels. Our EOS-token design achieves comparable WER and UTMOS without relying on oracle information, while maintaining stable generation speed, making it a practical choice for unified MLLM-based TTS.
% \subsection{Ablation3: EndPoint}
\begin{table}[h]
\centering
\caption{Performance with different stopping criteria. 
GT-Dur.: using oracle duration; 
GT-EP.: using oracle end-of-speech (oracle stopping point); 
EOS Token: our method, where the LM head predicts an end-of-sequence token during inference. }
% \resizebox{0.9\linewidth}{!}{%
\begin{tabular}{lcccccccc}
\toprule
\textbf{Stopping Criteria} & \textbf{WER (\%)$\downarrow$} & \textbf{SIM-R$\uparrow$}&  \textbf{SIM-G$\uparrow$} &\textbf{UTMOS$\uparrow$} \\
\midrule
GT-Dur.&29.36&0.48&0.43 & 2.55  \\
GT-EP.&3.46&0.49 &0.46   &3.21  \\
EOS Token& 3.61&0.49 &0.46   & 3.21  \\
\bottomrule
\end{tabular}
\label{tab:end_point}
% }
\end{table}

\textbf{Inference Hyperparameters.}
Table~\ref{tab:inference_param} analyzes the influence of diffusion inference parameters. Lower temperatures tend to produce cleaner but truncated outputs, leading to higher WER and lower similarity. Conversely, higher temperatures improve diversity but may reduce naturalness. We find that a temperature of 0.9 with 100 denoising steps achieves the best trade-off, yielding the lowest WER (1.95\%), highest similarity (0.54), and best UTMOS (4.00).

\begin{table}[h]
\centering
\caption{Performance under different inference parameters. }
% \resizebox{0.8\linewidth}{!}{%
\begin{tabular}{ccccccccc}
\toprule
\textbf{Temperature} &\textbf{Inference Steps} & \textbf{WER(\%)$\downarrow$} & \textbf{SIM-R$\uparrow$}&  \textbf{SIM-G$\uparrow$} &\textbf{UTMOS$\uparrow$} \\
\midrule
1&200&15.06&0.47&0.44 & 2.40 \\
1&100&7.53&0.48&0.44& 3.27 \\
0.9&100&1.95&0.54 &0.50&  4.00  \\
0.8&100&16.11&0.45& 0.41& 3.01  \\
0.8&80&19.88&0.44 & 0.39& 4.07 \\
\bottomrule
\end{tabular}
\label{tab:inference_param}
% }
\end{table}
% \subsection{Qualative result}
\section{Conclusion}
\label{sec:conclusion}

In this work, we present a dual-head multimodal language model that integrates a frame-level continuous-token diffusion head with an autoregressive LLM backbone for speaker-referenced TTS. By combining continuous speech representations, our approach avoids the quantization bottleneck and achieves high-fidelity and natural speech. To overcome exposure bias and improve training performance, we adapt masked training and a two-stage optimization scheme, which together substantially improve robustness and quality. Evaluations on LibriSpeech(PC) demonstrate significant gains, including a 46\% relative WER reduction over our baseline, along with higher speaker similarity and audio quality. These results highlight the effectiveness of bridging autoregressive modeling with diffusion-based refinement for continuous speech generation. 
Looking ahead, this framework provides a path toward unified foundation models that can support multiple speech and multimodal tasks within a single framework.

\bibliography{reference}

\end{document}